\documentclass[epj]{svjour}
\usepackage{graphics}
\usepackage{amsmath}
\usepackage{amsfonts}
\usepackage{amssymb}
\usepackage{graphicx}
\begin{document}

\title{Residue cross sections of $^{50}$Ti-induced fusion reactions based on the
two-step model}
\author{Ling Liu\inst{1}
\and Caiwan Shen\inst{2} \thanks{Corresponding author (email: cwshen@zjhu.edu.cn )}
\and Qingfeng Li\inst{2}
\and Ya Tu\inst{1}
\and Xiaobao Wang\inst{2}
\and Yongjia Wang\inst{2} }
\institute{College of Physics Science and Technology, Shenyang Normal University,
Shenyang 110034, Liaoning, China \and School of Science, Huzhou University,
Huzhou 313000, Zhejiang, China}
\date{Received: date / Revised version: date}

\abstract{
$^{50}$Ti-induced fusion reactions to synthesize superheavy elements
are studied systematically with the two-step model developed recently,
where fusion process is divided into approaching phase and formation
phase. Furthermore, the residue cross sections for different neutron
evaporation channels are evaluated with the statistical evaporation
model. In general, the calculated cross sections are much smaller
than that of $^{48}$Ca-induced fusion reactions, but the results are
within the detection capability of experimental facilities nowadays.
The maximum calculated residue cross section for producing superheavy
element $Z=119$ is in the reaction $^{50}$Ti+$^{247}$Bk in $3n$ channels
with $\sigma_{\rm res}(3n)=0.043$ pb at $E^{*}$ = 37.0 MeV.
}

\maketitle

\section{Introduction}

\label{intro} The synthesis of superheavy nuclei is a hot study field in
nuclear physics, and it has obtained much progress experimentally and
theoretically in recent years. Up to now, with the detecting of element $Z =
117$ in Dubna in 2010 \cite{Ref1}, the superheavy elements $Z = 110 - 118$
have been all synthesized \cite{Ref2,Ref3,Ref4}. Theoretical supports for
these very time-consuming and very high-expensive experiments are extremely
vital for choosing the suitable target-projectile combinations and the optimum
incident energy, and for the estimation of residue cross sections.

In synthesis of superheavy elements with proton number $Z=114-118$, the
so-called hot fusion reaction with $^{48}$Ca as a projectile and actinide as a
target is adopted. However, it comes increasingly difficult to synthesize
heavier elements with projectile $^{48}$Ca. Maybe the last superheavy element
which can be produced in the reaction with $^{48}$Ca is the element with $Z =
118$ since the target heavier than Cf is too difficult to be obtained. Thus to
produce heavier elements, heavier projectiles such as $^{50}$Ti, $^{54}$Cr,
$^{58}$Fe, $^{64}$Ni should be used.

Nevertheless, it is well known that in heavy ion induced reactions the
deep-inelastic and quasi-fission processes are the dominant reaction channels
because of the strong fusion hindrance, thus the fusion probability is much
smaller than the light ion induced reactions. In the reactions with $^{48}$Ca
and actinide targets the probability of fusion relative to quasi-fission is
less than $10\%$, and the ratio decreases for more symmetrical
target-projectile combinations \cite{Ref5}. Mass asymmetry is one of the
factors that influence quasi-fission and true fusion competition. Generally
speaking, a decrease in the mass asymmetry in the reaction entrance channel
leads to an increase in the quasi-fission and a decrease in the fusion
contributions into the capture cross sections. It appears that fusion is
strongly hindered as the size of the projectile relative to the target
increases. Therefore, it is not at all surprising to see the failed attempt to
make even heavier element $120$ using the $^{58}$Fe + $^{244}$Pu reaction
\cite{Ref6}. Thus, a spherical neutron magic nucleus $^{50}$Ti (only two
protons greater than Ca) seems to be a promising candidate of projectile in
the synthesis of superheavy element heavier than $Z = 118$. Furthermore, it is
worth mentioning that the lower limit of residue cross section which can be
detected experimentally is in the magnitude of 0.03 pb at present \cite{Ref7},
so synthesis of superheavy elements with projectile $^{50}$Ti is of great interest.

According to the theory of compound nucleus reactions, the whole process of
synthesizing the superheavy nuclei is composed of fusion part and fission
part. In the former part the projectile is captured by the target and a
amalgamated system is formed that then evolves into a spherical compound
nucleus, and then in latter part, besides being cracked into smaller
fragments, few of the compound nucleus may cool down by evaporating particles
and $\gamma$-rays and goes to its ground state. The evaporation residue cross
section is usually expressed as a sum over all partial waves at a certain
incident energy,
\begin{equation}
\label{eqn-1}\sigma_{\rm res}(E_{\rm c.m.})=\frac{\pi\hbar^{2}}{2\mu
E_{\rm c.m.}} \sum_{J}(2J+1)P_{\rm fus}^{J}(E_{\rm c.m.})\cdot
P_{\rm surv}^{J}(E_{\rm c.m.})
\end{equation}
where $J$ is the total angular momentum quantum number, $E_{\rm c.m.}$the
incident energy in the center of mass system; $P_{\rm fusion}^{J}$ and
$P_{\rm surv}^{J}$ denote the fusion and the survival probabilities, respectively.

In the evaporation process, though there is a certain margin of uncertainty in
the estimations of evaporation residue cross sections \cite{Ref8}, the
statistical evaporation model is commonly accepted and used to calculate the
evaporation probability $P_{\rm{surv}}^{J}$. However in the fusion
process, because of the complexity of heavy ion reactions, there is still no
commonly accepted model to deal with this process. Several models are adopted
to study the fusion reactions, such as fusion-by-diffusion model
\cite{Ref9,Ref10}, DNS model \cite{Ref11,Ref12}, QMD-based model \cite{Ref13},
etc. In this paper we adopt two-step model to study the $^{50}$Ti-induced
fusion reactions leading to the synthesis of superheavy nuclei.

The paper is arranged as follow: Sec. 2 gives a brief description of the
two-step model, and of the determination of parameters fitting the
experimental capture cross sections; Sec. 3 shows the results of systematic
calculations and discussions; Sec. 4 gives a summary.

\section{Two-step model and determination of parameters}

\label{sec:1}

The two-step model was proposed to describe the fusion process in massive
nuclear systems where fusion hindrance exists, as shown in Ref.
\cite{Ref14,Ref15,Ref16}. In this model, the fusion process is divided into
two stages: first, the sticking stage where projectile and target come to the
touching point over the Coulomb barrier from infinite distance, and second,
the formation stage where the touched projectile and target evolve to form a
spherical compound nucleus. Therefore, the fusion probability gets the form,
\begin{equation}
\label{eqn-2}P_{\rm{fusion}}^{J}(E_{\rm c.m.})=P_{\rm stick}
^{J}(E_{\rm c.m.}) \cdot P_{\rm{form}}^{J}(E_{\rm c.m.}).
\end{equation}
The energy dissipations in sticking stage and in formation stage are subtly
considered in the model. It is worth to emphasize that the two-step model
provides a method for a connection between a two-body collision process and
the subsequent one-body shape evolution. This is completely different from the
adiabatic, or the diabatic connection, and should be called as a ``statistical
connection'\ \cite{Ref14}.

In principle, both of the sticking probability and formation probability need
to be calculated via fluctuation-dissipation model, as shown in Ref.
\cite{Ref14}. However, for simplicity, we also choose alternatively an
empirical formula \cite{Ref17} to calculate the sticking probability, where
the barrier height is supposed to be Gaussian-distributed around the Coulomb
barrier to simulate the energy dissipation in the approaching phase. The
$P_{\rm stick}$ takes the form,
\begin{align}
P_{\rm stick}^{J}(E_{\rm c.m.})  &  =\frac{1}{2}\left\{  1+\operatorname{erf}\left[
\frac{1}{\sqrt{2}H}(E_{\rm c.m.}-B_{0}\right.  \right. \nonumber\\
&  -\left.  \left.  \frac{\hbar^{2}J(J+1)}{2\mu R_{B}^{2}})\right]  \right\}
, \label{eq3}
\end{align}
where $B_{0}$ is the barrier height of the Coulomb potentia, $H$ the width of
the Gaussian distribution of the barrier height, $\mu$ the reduced mass, and
$R_{B}$ the distance between two centers of projectile and target at the
Coulomb barrier. In a reasonable assumption $B_{0}/(\sqrt{2}H)$ should be much
greater than 1.

\begin{figure}[ptb]
\begin{center}
\resizebox{0.45\textwidth}{!}{\includegraphics{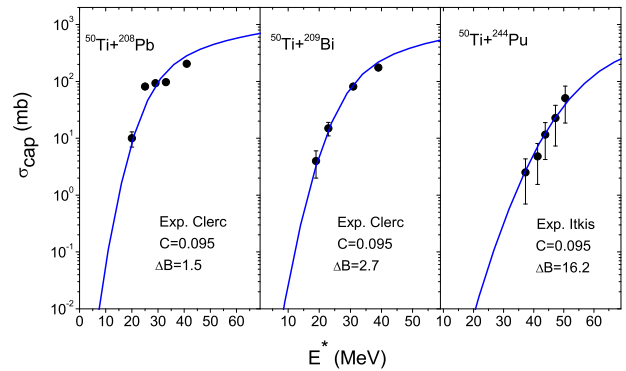}}
\end{center}
\caption{Fitting (solid lines) the experimental capture cross sections
\cite{Ref5,Ref18} to get appropriate $C$ and $\Delta B$.}
\label{fig:1}
\end{figure}

\begin{figure}[ptb]
\begin{center}
\resizebox{0.35\textwidth}{!}{\includegraphics{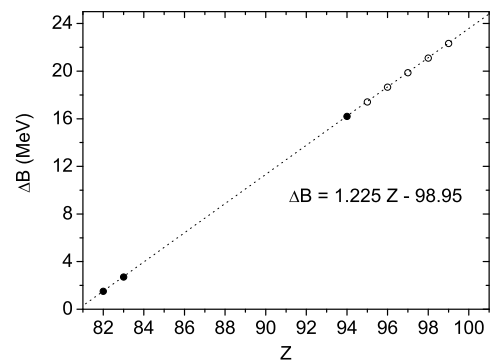}}
\end{center}
\caption{The extrapolation of the shift of Coulomb barrier for the nearby
heavier targets. The solid circles correspond to the data fitting to
experimental data in Fig. \ref{fig:1}, while the open ones are extrapolations
for Am, Cm, Bk, Cf, and Es.}
\label{fig:2}
\end{figure}

To calculate sticking probability, the parameter $C$ (a factor to calculate
the width of the Gaussian distribution of the barrier height, see Ref.
\cite{Ref15}) and the barrier height of the Coulomb potential $B_{0}=B+\Delta
B$ should be adjusted as adequate as possible for very heavy systems. The
three systems to be fitted are $^{50}$Ti+$^{208}$Pb, $^{50}$Ti+$^{209}$Bi
\cite{Ref18} and $^{50}$Ti+$^{244}$Pu \cite{Ref5}. The fitted results are
shown in Fig. \ref{fig:1}. With a constant value of $C=0.095$ and a linear
increase of the barrier height shift $\Delta B$ with proton number $Z$, the
available experimental capture cross sections are well reproduced. The $\Delta
B$ is extrapolated from the fitting formula $\Delta B=1.225Z-98.95$ for heavy
targets having very close atomic numbers, namely Am, Cm, Bk, Cf, and Es, as is
seen in Fig. \ref{fig:2}.

With Eq. (\ref{eq3}) and the re-fitted parameter $C$ and $\Delta B$, the sticking
probability of $^{50}$Ti-induced fusion reactions is calculated with
confidence. Further more, for hot fusion reactions, we are interested in the
residue cross sections around $E^{\ast}=30\sim40$ MeV, and the experimental
data are just located in the similar energy range, as shown in Fig.
\ref{fig:1}, the calculated $\sigma_{\rm{res}}$ around the same energy
range will not sensitively depends on the determination of $C$ and $\Delta B$.

\section{Residue cross sections}

\label{sec:2}

In the excited compound nucleus, the de-excitation process includes usually
light particle emissions, $\gamma$-ray emissions, and fission. However,
because of existence of the Coulomb barrier for charged particle emissions,
the probability for the emission of light charged particles is much smaller
than the one for the neutron emission. Therefore, most of the superheavy
nuclei are obtained through the consecutive neutron evaporations. In the
calculation of survival probability, the \textsc{hivap} code, based on the
statistical evaporation model, is adopted to evaluate the residue cross sections.

The very important parameter in the evaporation process is the fission barrier
$B_{f}$. It is clear that for heavy nuclei, the more stable one, which means
having larger shell correction energy $E_{\rm{shell}}$, usually has a
heigher fission barrier. Thus, the classical way to calculate the fission
barrier is $B_{f}=B_{\rm{LD}}-E_{\rm{shell}}$, as did in Ref.
\cite{schmidt1991}. However the microscopic calculations does not prove such
so simple relationship between $B_{f}$ and $E_{\rm{shell}}$. Since the
microscopic calculations do not give $B_{f}$ for so heavy compound nuclei
discussed in the present paper, we thus used the classical form to calculate
$B_{f}$, but with an arbitrary factor $f$ to the shell correction energy,
i.e.,
\begin{equation}
B_{f}=B_{\rm{LD}}-f\cdot E_{\rm{shell}},
\end{equation}
where $E_{\rm{shell}}$ is taken from Ref. \cite{Ref19}, and the factor $f$
is determined by fitting the experimental data. In the present case, the
inclusion of the factor $f$ gives rise to reductions of the fission barrier,
and then the reduction of residue cross sections, but does not change general
feature of the excitation functions, i.e., peak positions etc., though
decreasing slopes in higher energies are a little affected. The introduction
of the factor $f$, thus, is appropriate for predictions of residue cross sections.

Up to now, there are no experimental residue cross sections for $^{50}
$Ti-induced fusion reactions to synthesize superheavy nuclei with $Z\geq110$.
Fortunately, the residue cross sections of $^{48}$Ca+$^{249}$Bk had been
measured by Dubna in 2010 \cite{Ref1}, and together with fitting the
experimental residue cross section of $^{48}$Ca+$^{208}$Pb \cite{Ref20} and
$^{50}$Ti+$^{208}$Pb \cite{Ref18}, the corresponding factor $f$ for the three
systems are determined to be $0.45$(which is the same as in Ref.
\cite{Ref15}), $0.72$ and $0.77$, respectively, using HIVAP code. Then, the
factor $f$ for reaction $^{50}$Ti +$^{249}$Bk can be approximately evaluated
as $0.45\times(0.77/0.72)=0.48$. Since the target Bk of the reaction have only
one or two protons more or less than the targets such as Am, Cm, Cf and Es,
the factor $f$ for Ti+Bk should also work with these target in $^{50}
$Ti-induced reactions.

\begin{figure}[ptb]
\begin{center}
\resizebox{0.4\textwidth}{!}{\includegraphics{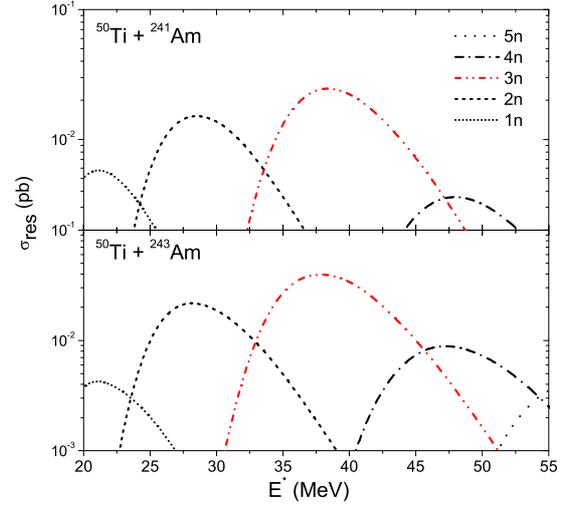}}
\end{center}
\caption{Predicted residue cross sections for producing superheavy element
$Z=117$. The short dot line, short dash line, dash dot dot line, dash dot
line, and dot line represent for $1n$, $2n$, $3n$, $4n$, and $5n$ neutron
evaporation channels, respectively.}
\label{fig:3}
\end{figure}

\begin{figure}[ptb]
\resizebox{0.5\textwidth}{!}{\includegraphics{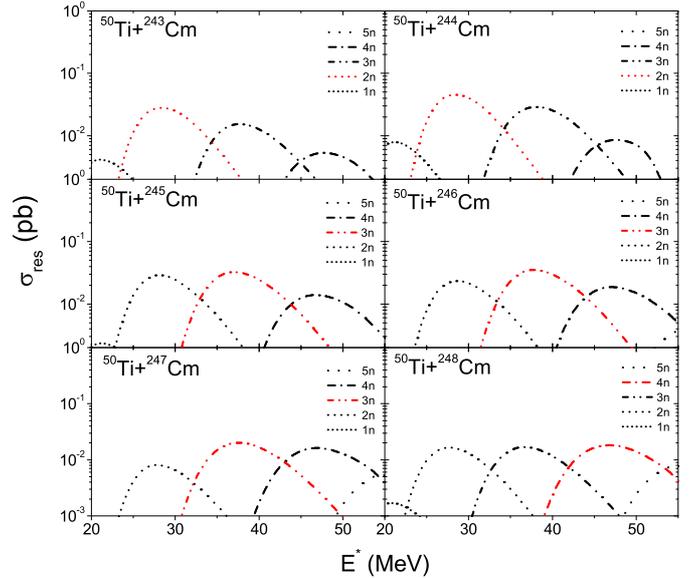}
}\caption{The same as in Fig. 3, but for producing superheavy element $Z=118$.
}
\label{fig:4}
\end{figure}

\begin{figure}[ptb]
\begin{center}
\resizebox{0.4\textwidth}{!}{\includegraphics{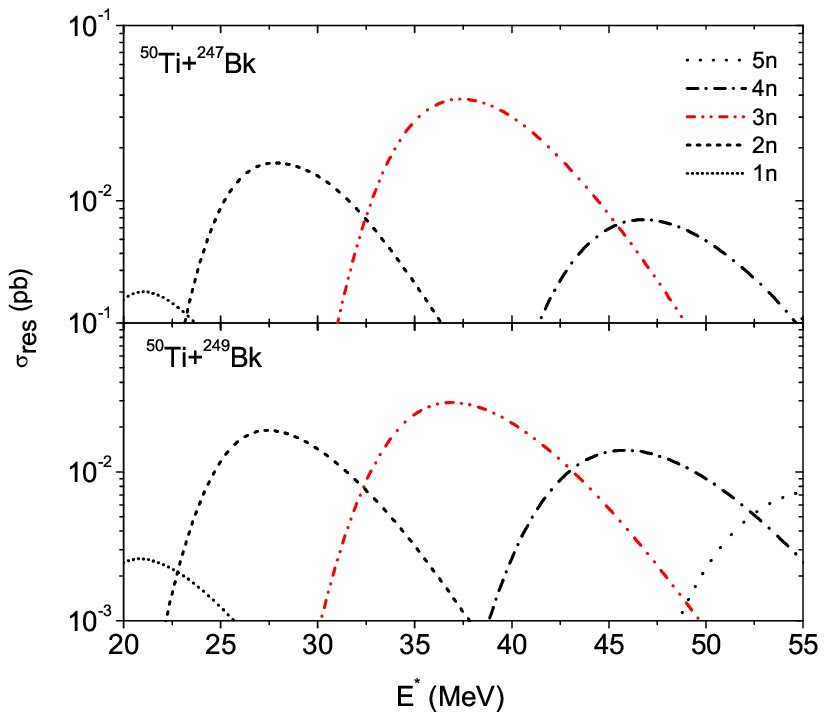}}
\end{center}
\caption{The same as in Fig. \ref{fig:3}, but for producing superheavy element
$Z=119$.}
\label{fig:5}
\end{figure}

\begin{figure}[ptb]
\begin{center}
\resizebox{0.48\textwidth}{!}{\includegraphics{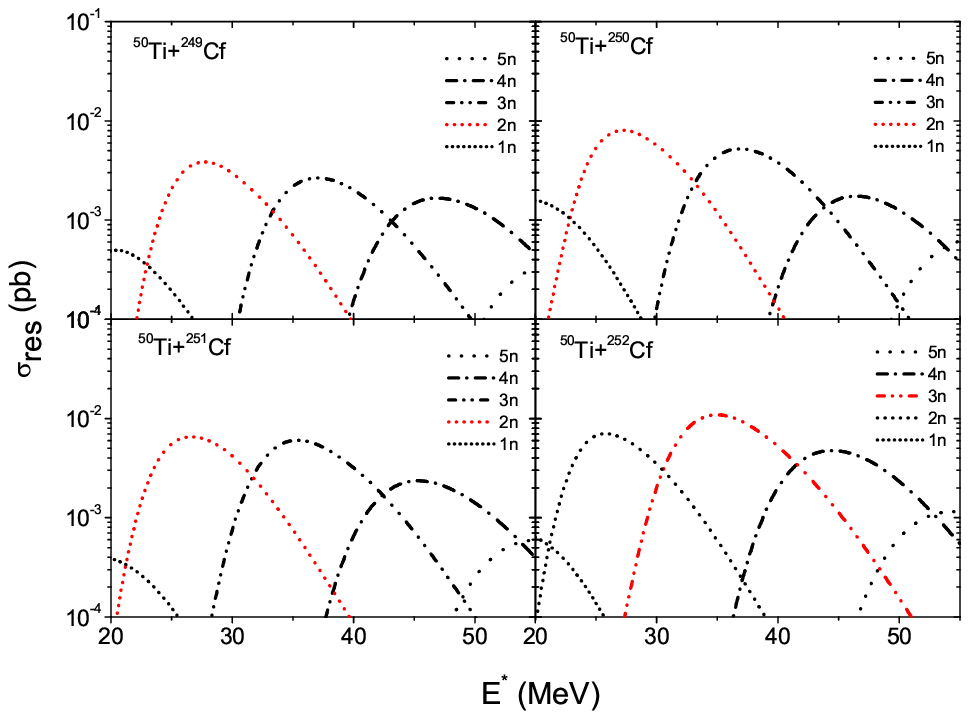}}
\end{center}
\caption{The same as in Fig. \ref{fig:3}, but for producing superheavy element
$Z=120$.}
\label{fig:6}
\end{figure}

\begin{figure}[ptb]
\begin{center}
\resizebox{0.40\textwidth}{!}{\includegraphics{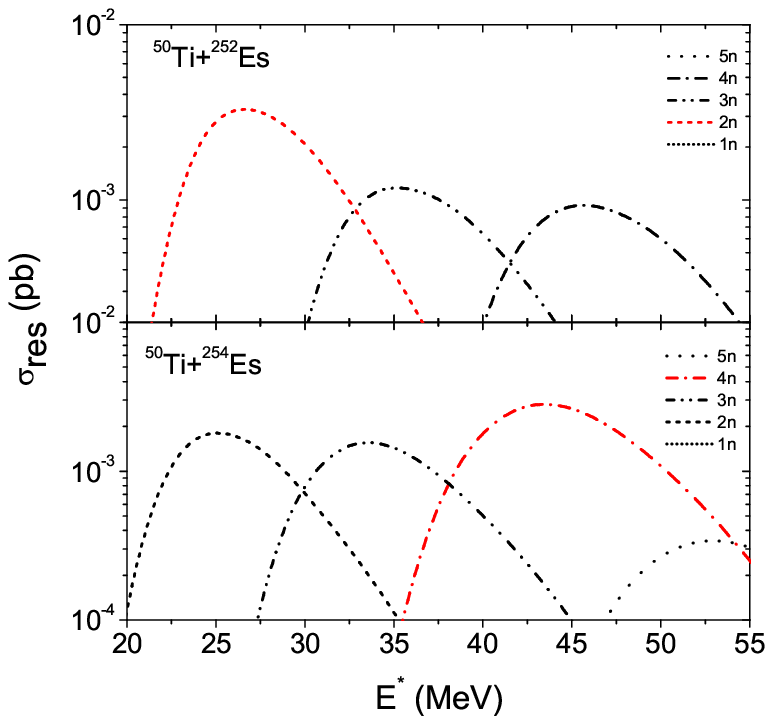}}
\end{center}
\caption{The same as in Fig. \ref{fig:3}, but for producing superheavy element
$Z=121$.}
\label{fig:7}
\end{figure}

Now, all the ingredients in our calculations do not leave any ambiguity. The
target isotopes of Am, Cm, Bk, Cf, and Es with life times long enough for
experiments are chosen, and then, the evaporation residue cross sections with
$Z=117-121$ in some selected reactions are calculated systematically using the
two-step model and HIVAP with the purpose of searching the favorable reaction
systems and collision energies for the synthesis of superheavy nuclei with
even larger proton number. The corresponding results are shown in Figs.
\ref{fig:3}-\ref{fig:7}.

Fig. \ref{fig:3} shows very similar excitation functions of producing element
$Z=117$ in the $^{50}$Ti+$^{241,243}$Am reactions. It is reasonable since two
targets differ with only two neutrons. The maximum residue cross sections for
$^{50}$Ti+$^{243}$Am, is slightly larger with at in $3n$ channels.

The calculated residue cross sections for reactions $^{50}$Ti+ $^{243-248}$Cm
to produce elements $Z=118$ are presented in Fig. \ref{fig:4}. According to
the results, the optimum reaction to synthesize $Z=118$ are $^{50}$Ti+
$^{244}$Cm with $\sigma_{\rm{res}}(2n)= 0.053$ pb at $E^{*}= 28.0$ MeV and
$^{50}$Ti+ $^{246}$Cm with $\sigma_{\rm{res}} (3n)$ = 0.040 pb at $E^{*}=
37.1$ MeV. As one expects, the variation of the peak energies between $2n$ and
$3n$ evaporation channels of the compound nuclei is around 8 MeV which is
about one neutron separation energy. Furthermore, it can be seen from Fig.
\ref{fig:4} that our results do not show strong isotope dependence of
superheavy nucleus production. Generally speaking, the formation of the
superheavy nucleus is a complex dynamical process and depends on many physical
factors, such as Coulomb barrier, conditional saddle point, neutron separation
energy, shell effect and so on. It needs more further explorations.

The next superheavy element to be synthesized in experiment may be $Z = 119$
or $120$, therefore, the investigations of the synthesis of $Z > 118$,
especially of $119$ and $120$, are interesting and useful. It shows in Fig.
\ref{fig:5} that in reactions $^{50}$Ti+$^{247,249}$Bk the maximum residue
cross sections for producing superheavy element $Z=119$ are both in $3n$
channels and are, respectively, 0.043 pb and 0.033 pb, which are almost one
order of magnitude smaller than that of $^{48}$Ca+$^{247,249}$Bk in reference
\cite{Ref15}. This could be explained by the fact that for the same actinide
target, the Coulomb potential for $^{50}$Ti-induced reaction is roughly $10\%$
larger than that for $^{48}$Ca-induced reactions, and the fusion hindrance for
the former one is also stronger than the latter case \cite{Ref21}. Moreover,
the results for $^{50}$Ti+$^{249}$Bk from Ref.\cite{Ref9} with
fusion-by-diffusion model and from Ref.\cite{Ref22} with dinuclear system
model are 0.57 pb and 0.11 pb respectively, while our present calculation
value is very close to 0.035 pb calculated with an analytical expression for
the description of the fusion probability \cite{Ref23} and 0.05 pb in Ref.
\cite{Ref24}. The different results can be attributed to the dependence of
model and parameter. However, although the cross sections are relatively small
and are more than two orders of magnitudes lower than pico-barn, they are in
the detection capability of the present experimental facilities.

Fig. \ref{fig:6} and Fig. \ref{fig:7} gives the results for producing $Z =
120$ and $121$ in reactions $^{50}$Ti + $^{249-252}$Cf and $^{50}$Ti +
$^{252,254}$Es, respectively. It can be seen from the figures that the residue
cross sections are too small, approaching to the order of femto-barn. As was
noted in the figures, the maximum residue cross sections in the reactions
$^{50}$Ti + $^{250, 251, 252}$Cf are, 9.7 fb, 7.5 fb, and 12.2 fb,
respectively, which are obviously larger than those of the reaction $^{50}$Ti
+ $^{249}$Cf (4.6 fb). Recently, Siwek-Wilczynska et al. predict the cross
section of $^{50}$Ti + $^{249}$Cf to synthesize the element $Z=120$ to be only
6 fb \cite{Ref10} which is more consistent with our present result. In
contrast, our result is smaller than other several predictions with
$\sigma_{\rm{res}}\approx20\sim200$ fb \cite{Ref9,Ref23,Ref24}. In
addition, it is worth arguing that the residue cross section with target
$^{252}$Cf is several times larger than those of the targets $^{249-251}$Cf
and hence is theoretically the most favorable one for the synthesis of element
$120$. However, $^{252}$Cf may be difficult to be target because its
spontaneous fission would bring about serious background in the experiment.
Therefore, to produce element $Z=120$, $^{50}$Ti + $^{249-251}$Cf could be
considered in the future experiments. As a further extension, we evaluate the
residue cross sections of element $121$ with targets $^{252, 254}$Es though
Einsteinium is rather exotic and may be hardly prepared presently. It shows
that the maximum residue cross sections for nucleus $Z=121$ have comparable
value of about 3 fb and are far below the present experimental limit of
registration (30 fb). Thus, the synthesis of elements with $Z>120$ is rather
problematic in the near future due to extremely low cross sections and short
half-lives of these elements.

To illustrate the results clearly, the relatively larger residue cross
sections for different reactions are listed in Table \ref{tab:1}. It should be
mentioned that our reduction factor $f$ of $E_{\rm{shell}}$ of the
compound system influence only on the absolute values of residue cross
sections, not on the shapes of the residue excitation functions.
\begin{table}[ptb]
\caption{The relatively larger residue cross sections for the $^{50}
$Ti-induced reactions to synthesize superheavy nuclei for different target
elements. The half-lives of the targets are taken from Ref. \cite{halflife}.}
\label{tab:1}
\begin{center}
\begin{tabular}
[c]{ccccc}\hline\hline
$Z_{CN}$ & Target & $T_{1/2}$(target) & $E^{\ast}$(MeV) & $\sigma
_{\rm{res}}(pb)$\\\hline
117 & $^{243}$Am & 7370 y & 37.3 & 0.044 (3n)\\
118 & $^{244}$Cm & 18.10 y & 28.0 & 0.053 (2n)\\
118 & $^{246}$Cm & 4730 y & 37.1 & 0.040 (3n)\\
119 & $^{247}$Bk & 1380 y & 37.0 & 0.043 (3n)\\
120 & $^{252}$Cf & 2.645 y & 35.3 & 0.012 (3n)\\
121 & $^{252}$Es & 471.7 d & 26.5 & 0.004 (2n)\\\hline\hline
\end{tabular}
\end{center}
\end{table}

\section{Summary}

\label{sec:3} In summary, $^{50}$Ti-induced fusion reactions to synthesize
superheavy nucleus with $Z=117\sim121$ are studied with the two-step model and
statistical evaporation model, where fusion process is divided into two
consecutive phases, i.e., approaching phase and formation phase. The results
show that the reactions $^{50}$Ti + $^{241,243}$Am, $^{50}$Ti + $^{243-248}
$Cm, $^{50}$Ti + $^{247,249}$Bk to synthesize superheavy nucleus with $Z=117$,
$118$ and $119$ have smaller residue cross sections than $^{48}$Ca-induced
ones with nearly one order of magnitude. However, the calculated residue cross
sections are still within the detection capability of experiment nowadays.
Whereas, $^{50}$Ti-induced fusion reactions with a target $^{249-252}$Cf, and
$^{252,254}$Es, to synthesize superheavy elements with $Z=120$ and $121$
respectively, have so small residue cross sections that the experiments can be
performed only when the experimental facilities are developed in the future.
Of course, for planning the experiments on the synthesis of superheavy nuclei
of up to $Z = 122$, new mechanism and more precise data obtained in the
processes of fusion-fission and quasi-fission of these nuclei are required.

\section{Acknowledgements}

\label{sec:4} This work was supported by the National Natural Science
Foundation of China (Grant Nos. 11275068, 11547312, 11375062, 11505057,
11505056 and 11305108), and the project sponsored by SRF for ROCS, SEM. One of
the authors (L.L) is grateful to the hospitality and calculation support of
C3S2 Computing Center by Huzhou University. The author (C.W.Shen) acknowledges
the fruitful discussions and suggestions from Prof. Y. Abe and D. Boilley, and
the hospitality by RCNP Osaka University and GANIL.


\begin{thebibliography}{99}

\bibitem {Ref1}Y. T. Oganessian, F. S. Abdullin, P. D. Bailey et al., Phys.
Rev. Lett. $\mathbf{104}$, (2010) 142502.

\bibitem {Ref2}Y. T. Oganessian, V. K. Utyonkov, Y. V. Lobanov et al., Phys.
Rev. C $\mathbf{62}$, (2000) 041604.

\bibitem {Ref3}Y. T. Oganessian, V. K. Utyonkov, Y. V. Lobanov et al., Phys.
Rev. C $\mathbf{69}$, (2004) 021601.

\bibitem {Ref4}Y. T. Oganessian, V. K. Utyonkov, Y. V. Lobanov et al., Phys.
Rev. C $\mathbf{74}$, (2006) 044602.

\bibitem {Ref5}M. G. Itkis, A. A. Bogachev, I. M. Itkis et al., Nucl. Phys. A
$\mathbf{787}$, (2007) 150c--159c.

\bibitem {Ref6}Y. T. Oganessian, V. K. Utyonkov, Y. V. Lobanov et al., Phys.
Rev. C $\mathbf{79}$, (2009) 024603.

\bibitem {Ref7}K. Morita et al., J. Phys. Soc. Jpn. $\mathbf{76}$, (2007)
043201; Morita K, et al., J. Phys. Soc. Jpn. $\mathbf{76}$, (2007) 045001.

\bibitem {Ref8}K. Swiwek-Wilczynska, I. Skwira, Phys. Rev. C $\mathbf{72}$,
(2005) 034605.

\bibitem {Ref9}Z. H. Liu, J. D. Bao, Phys. Rev. C $\mathbf{84}$, (2011) 031602(R).

\bibitem {Ref10}K. Siwek-Wilczynska, T. Cap, M. Kowal et al., Phys. Rev. C
$\mathbf{86}$, (2012) 014611.

\bibitem {Ref11}W. Li, N. Wang, J. F. Li et al., Euro. Phys. Lett.
$\mathbf{64}$, (2003) 750.

\bibitem {Ref12}Z. Q. Feng, G. M. Jin, J. Q. Li et al., Phys. Rev. C
$\mathbf{76}$, (2007) 044606.

\bibitem {Ref13}N. Wang, X. Z. Wu, Z. X. Li et al., Phys. Rev. C $\mathbf{74}
$, (2006) 044604.

\bibitem {Ref14}C. W. Shen, G. Kosenko, Y. Abe, Phys. Rev. C $\mathbf{66}$,
(2002) 061602.

\bibitem {Ref15}C. W. Shen, Y. Abe, D. Boilley et al., Int. J. Mod. Phys. E
$\mathbf{17}$(Suppl), (2008) 66--79.

\bibitem {Ref16}C. W. Shen, Y. Abe, Q. F. Li et al., Sci. China. Ser. G-Phys.
Mech. Astron. $\mathbf{10}$, (2009) 1458--1463.

\bibitem {Ref17}W. J. Swiatecki, K. Siwek-Wilczynska, J. Wilczynski, Phys.
Rev. C $\mathbf{71}$, (2005) 014602.

\bibitem {Ref18}H. G. Clerc, J. G. Keller, C. C. Sahm et al., Nucl. Phys. A
$\mathbf{419}$, (1984) 571-588.

\bibitem {schmidt1991}K-H Schmidt and W. Morawek, Rep. Prog. Phys.
\textbf{54}, (1991) 949-1003.

\bibitem {Ref19}P. M\"{o}ller, J. R. Nix, W. D. Myers et al., At. Data. Nucl.
Data. Tables. $\mathbf{59}$, (1995) 185--381.

\bibitem {Ref20}Y. T. Oganessian, V. K. Utyonkov, Y. V. Lobanov et al., Phys.
Rev. C $\mathbf{64}$, (2001) 054606.

\bibitem {Ref21}C. W. Shen, D. Boilley, Q. F. Li, J. J. Shen et al., Phys.
Rev. C $\mathbf{83}$, (2011) 054620.

\bibitem {Ref22}N. Wang, E. G. Zhao, W. Scheid et al., Phys. Rev. C
$\mathbf{85}$, (2012) 041601.

\bibitem {Ref23}N. Wang, J. L. Tian, W. Scheid, Phys. Rev. C $\mathbf{84}$,
(2011) 061601(R).

\bibitem {Ref24}v. I. Zagrebaev, W. Greiner, Phys. Rev. C $\mathbf{78}$,
(2008) 034610.

\bibitem {halflife}G. Audi, F. G. Kondev, M. Wang, B. Pfeiffer, X. Sun, J.
Blachot, and M. MacCormick, Chinese Phys. C 36, (2012) 1157.
\end{thebibliography}
\end{document}